\documentclass[prc,twocolumn,nofootinbib,showpacs,floatfix,letterpaper]{revtex4}

\usepackage{graphicx}
\usepackage{amsmath,amssymb}
\usepackage{multirow,longtable,color}
\usepackage[body={7.33in,9.45in},top=0.75in]{geometry}
\newcommand{\eq}[1]{\begin{equation} #1 \end{equation}}

\newcommand{\overlap}[2]{\langle #1 | #2 \rangle}
\newcommand{\elmx}[3]{\langle #1|#2|#3 \rangle}
\newcommand{\ket}[1]{|#1\rangle}
\newcommand{\bra}[1]{\langle #1|}
\newcommand{\ac}[1]{a_{#1}^{\dag}}
\newcommand{\Tsqr}{\hat{\mathbf{T}}^2}

\newcommand{\tsqr}{\hat{\mathbf{t}}^2}
\newcommand{\tz}{\hat{t}_z}
\newcommand{\tp}{\hat{t}_+}
\newcommand{\tm}{\hat{t}_-}

\renewcommand{\arraystretch}{1.2}

\begin{document}


\title{Isospin mixing in a particle-number conserving microscopic 
approach}

\author{L. Bonneau\textsuperscript{1}, J. Bartel\textsuperscript{2} 
and P. Quentin\textsuperscript{1,3}}

\affiliation{
\textsuperscript{1}Theoretical Division, Los Alamos
National Laboratory, Los Alamos, NM 87545, USA \\ 
\textsuperscript{2}Institut Pluridisciplinaire Hubert Curien, 
CNRS/IN2P3 and Universit\'e Louis Pasteur 
67000 Strasbourg, France \\
\textsuperscript{3}Centre d'Etudes Nucl\'eaires de Bordeaux-Gradignan, 
Universit\'e Bordeaux-I and CNRS/IN2P3, BP 120, 33175 Gradignan, France\\
}

\date{\today}

\begin{abstract}

We calculate the isospin-mixing parameter for several
$T_z=-1$, $T_z=0$ and $T_z=1$ nuclei from Mg to Sn in the
particle-number conserving  Higher Tamm--Dancoff approach taking into
account the pairing correlations. In particular we investigate the
role of the Coulomb interaction and the $|T_z|=1$ pairing correlations. To
do so the HTDA approach is implemented with the SIII Skyrme effective 
nucleon-nucleon
interaction in the mean-field channel and a delta interaction in the
pairing channel. We conclude from this investigation that the pairing
correlations bring a large contribution to isospin-symmetry
breaking, whereas the Coulomb interaction turns out to play a less
important role. Moreover we find that the isospin-mixing parameters for
$T_z=-1$ and $T_z=1$ nuclei are comparable while they are about twice as
large for $T_z=0$ nuclei (between 3\% and 6\%, including doubly magic 
nuclei).

\end{abstract}

\pacs{
{21.10.Hw}
{21.60.Jz}
{21.60.Cs}
}

\maketitle


\section{Introduction}

One of the most striking aspect of the structure of atomic nuclei is
the very small violation of the  isobaric invariance. This is so even
for heavy nuclei where the Coulomb interaction is not thought a priori
to act merely in a perturbative manner. As pointed out for instance in
Ref.~\cite{Bohr-Mottelson}, this is due to the weak variation of the
symmetry-breaking Coulomb field over the nuclear volume. It has also
been suggested from phenomenological and fundamental (at the
levels of quarks having different masses) points of view that genuine
isospin non-conserving parts of the strong interaction should be
considered. They should however be rather small as compared to their
conserving counterparts. 

As a consequence, it has been found that a nuclear ground state 
$\ket{\Psi}$ may be thought as being composed of mostly a $T_0=|T_z|$ 
component where $T_z=(N-Z)/2$ with a small $T_0+1$ admixture, namely 
\eq{
\label{alpha}
\ket{\Psi}\approx \beta \ket{T_0\,T_z} + \alpha \ket{T_0+1\,T_z}\:,
}
where $\alpha^2+\beta^2=1$.

Even though in most cases the isobaric invariance may be flatly
assumed, there are phenomena where a specific knowledge of the
isospin mixing is needed. This is in particular the case whenever some
observed transition or reaction would be forbidden, should this
invariance be exactly fulfilled. Interesting cases, where the isospin
mixing has to be considered, are also related with beta-decay properties 
(see, e.g., the review of Ref.~\cite{BlinStoyle}). Of particular importance
in that respect are the studies of superallowed $0^+$-to-$0^+$ nuclear
$\beta$ decays in the context of the tests of the CVC hypothesis (see,
e.g., Ref.~\cite{Hardy-Towner}) through $ft$-value
measurements. Hence, a specific determination of the effect of the
isospin mixing is required to correct the value yielded by the crude
isospin-multiplet approximation (determining thus the so-called
$\delta_{C}$ corrective term). 

Before entering, in a subsequent study, into a detailed assessment of
the transition matrix element involved in such particular decays, we
consider it interesting to evaluate first the actual importance of the
isospin mixing as measured for instance by $\alpha^2$. This is the
subject of the present paper. 

Presently available theoretical estimates of the isospin mixing fall into 
three different categories. 

First, one has to quote the hydrodynamical approach of Bohr,
Damg{\aa}rd and Mottelson~\cite{BDM} which consists in quantifying the
normal modes associated with the polarization effect of the Coulomb
field on a spherically symmetrical isovector density. In $N=Z$ nuclei
this approach yields the probability $\alpha^2$ of the
$T=1$ component, in sole addition to the dominating $T=0$ component,
which is given by
\eq{
\label{bohrmottelson}
\alpha^2 = 3.5\times10^{-7}Z^2 A^{2/3}\:.
}
It therefore amounts, e.g., for the $^{40}$Ca nucleus to about
0.16\%. In nuclei having a neutron excess, these authors estimate that
$\alpha^2$ (meaning now the probability of the $|T_z| + 1$ component
over the dominating $|T_z|$ component) is equal to the value given by
Eq.~(\ref{bohrmottelson}) divided by $|T_z| + 1$. This reduction, 
which is expressed in terms of a factor being merely the
square of a Clebsch--Gordan coefficient, has been first advocated by
Lane and Soper~\cite{Lane-Soper}. It yields, e.g., for the $^{48}$Ca
nucleus, a value of $\alpha^2$ of about 0.04\%.

The second class of approaches are based on shell-model
calculations. Their success is contingent, as usual within such an
approach, upon the relevance of the matrix elements in use. For the
description of isospin mixing, an accurate determination of Coulomb
matrix elements is of course of paramount importance (see for instance
the discussion of Coulomb energy differences in $A=47$ and $A=49$
mirror pairs~\cite{bentley}). This constitutes an a priori necessary
condition to provide valuable answers to the question left open on the
real importance of isospin non-conserving forces as studied for
example to explain the isobaric multiplet yrast energies in
Ref.~\cite{zuker}. Other concerns are related to a good description of
radial single-particle wave functions as in, e.g., Ref.~\cite{ormand}
to describe asymmetry factors in parity-violating electron
scattering. One definite difficulty of shell-model calculations is due
to the fact that they do not take into account any core isospin mixing, 
excepted of course for the no-core shell model calculations limited
to very light nuclei (see for instance Ref.~\cite{Navratil}).

One might then be inclined to think that microscopic calculations
making use of phenomenological nucleon-nucleon forces should be able
to describe the polarization effects of the Coulomb interaction at
least at the mean field level, in a satisfactory way. Indeed, as
opposed to shell-model calculations, mean-field calculations are
expected to provide rather elaborate single-particle wave functions and
they do not rely on any inert core approximation. However, apart from
possible consequences of well-known symmetry violations inherent to
the mean field approximation, they request as a next very important
step to account accurately for the correlations. This may be done
without serious a priori problems for RPA-type correlations, as
performed for instance in~\cite{sagawa,vangiai}. In
Ref.~\cite{sagawa}, it is shown that the hydrodynamical ansatz of
Ref.~\cite{BDM} underestimates the isospin mixing by a factor 2 to 4
(see Fig.~3 of~\cite{sagawa}). It is important to note that the latter
approach does not include important correlations, namely pairing
correlations. There are good practical reasons for such an
omission. The usual handling of pairing correlations within a kind of
Bogoliubov quasiparticle vacuum approximation as in the BCS or
Hartree--Fock--Bogoliubov theory is totally unfit for the isospin mixing
problem. Indeed, such an ansatz yields spurious components of both
charge state particle numbers, giving rise in turn to a spurious
mixing of $T_z$-components which invalidates a priori any attempt to
extract out of them any meaningful $T$-mixing properties.

This is why we make use here of the Higher Tamm--Dancoff approach
(HTDA) which can be interpreted as a highly truncated shell model
built on a self-consistent Hartree--Fock
solution~\cite{Pillet02,Pillet05,Sieja07,Bonneau07_N=Z}. At this stage
we focus on the role of $|T_z| =1 $ pairing correlations, which gives
us an upper limit of the isospin mixing parameter since proton-neutron
pairing correlations are expected to reduce the isospin mixing as it
will be discussed below. For the time-being we will not evaluate the
effect of RPA correlations which could be (and will soon be) easily
taken into account into the HTDA framework.

To determine $\alpha^2$, we should in principle perform a projection
of the ground state $\ket{\Psi}$ on good isospin states
$\ket{T\,T_z}$. Assuming, however, that components higher than $T_0+1$
are negligible, as in Eq.~(\ref{alpha}), we can deduce $\alpha^2$ from
the calculation of the expectation value of the square of the isospin
operator $\hat{\mathbf{T}}$ in the state $\ket{\Psi}$. Indeed, if
$\ket{\Psi}$ is normalized to unity and assuming that the dominant
contribution of the ground-state (GS) isospin comes from $T_0=|T_z|$, we have 
\eq{
\label{T2_general}
\begin{aligned}
\elmx{\Psi}{\Tsqr}{\Psi}= \left(1-\alpha^2\right)\,
T_0\left(T_0+1\right) +\alpha^2\,\left(T_0+1\right)\left(T_0+2\right)\:,
\end{aligned}
}
hence
\eq{
\label{alpha2}
\alpha^2=\frac{\elmx{\Psi}{\Tsqr}{\Psi}-T_0(T_0+1)}{2(T_0+1)}\:.
}

The paper is organized as follows. 
After the derivation of the expression for the expectation value of
$\Tsqr$ in the state $\ket{\Psi}$ in Sect.~\ref{expvalT2}, we present
in Sect.~\ref{res} the results of the HTDA calculations for the
GS properties and the values of all relevant isospin
quantities, such as the expectation value
$\elmx{\Psi}{\hat{\mathbf{T}}^2}{\Psi}$, the deduced value of $T$ and
the isospin-mixing parameter $\alpha^2$ for a large sample of
nuclei. The main conclusions of this study are drawn in
Sect.~\ref{concl}.


\section{Expectation value of $\Tsqr$ in the Higher Tamm--Dancoff
  approach}
\label{expvalT2}

%
%

\subsection{Correlated ground state in the Higher Tamm--Dancoff approach}

Neglecting here the proton-neutron residual interaction, we can write 
the many-body state $\ket{\Psi}$ describing the ground state of a
nucleus as the product of the correlated states $\ket{\Psi^{(n)}}$ and
$\ket{\Psi^{(p)}}$ 
\eq{
\label{Psi_HTDA}
\ket{\Psi}=\ket{\Psi^{(n)}}\otimes\ket{\Psi^{(p)}}\:,
}
where, in the HTDA approach, $\ket{\Psi^{(q)}}$ ($q=n$ for neutrons
and $q=p$ for protons) is a superposition of $N_q$-particle Slater
determinants ($N_q=N$ for neutrons and $N_q=Z$ for protons) of the form
\eq{
\label{Psi_tau_HTDA}
\ket{\Psi^{(q)}} = \chi_0^{(q)}\,\ket{\Phi_0^{(q)}} +
\sum_{i>0} \chi_i^{(q)} \,\ket{\Phi_i^{(q)}}\:.
}
In Eq.~(\ref{Psi_tau_HTDA}), $\ket{\Phi_0^{(q)}}$
denotes the Hartree--Fock (HF) ground state and the 
$\ket{\Phi_i^{(q)}}$ are $n$-particle--$n$-hole excited states
built on $\ket{\Phi_0^{(q)}}$\footnote{For 
the sake of clarity in the notation, we reserve the letter $\Phi$ for 
a Slater determinant and the letter $\Psi$ for a correlated
state.}. The a priori complex coefficients
$\chi_0^{(q)}$ and $\chi_i^{(q)}$ are determined by minimizing
the energy functional calculated for $\ket{\Psi^{(q)}}$. In fact,
in order for the many-body state $\ket{\Psi}$, when constructed with real
single particle wave functions, to be time-reversal
invariant, the coefficients $\chi_0^{(q)}$ and $\chi_n^{(q)}$
must be real.

%
%

\subsection{Expression of the expectation value of $\Tsqr$}

Since $\Tsqr$ is an hermitian operator (see Appendix~A for its
definition and properties), its expectation value in the
HTDA state $\ket{\Psi}$ reads
\eq{
\label{expval_Tsqr}
\begin{aligned}
& \elmx{\Psi}{\Tsqr}{\Psi} = \\
& \sum_{i,j} \bigl(\chi_i^{(n)}\chi_j^{(p)}\bigr)^2
\bigl(\bra{\Phi_i^{(n)}}\otimes\bra{\Phi_j^{(p)}}\bigr) \Tsqr
\bigl(\ket{\Phi_i^{(n)}}\otimes\ket{\Phi_j^{(p)}}\bigr)\\
& +2\sum_{\substack{i\leqslant i',j\leqslant j' \\ (i',j')\ne (i,j)}}
\chi_i^{(n)}\chi_j^{(p)}\chi_{i'}^{(n)}\chi_{j'}^{(p)}\, \times \\
& \phantom{+2}
\mathrm{Re}\biggl[\bigl(\bra{\Phi_i^{(n)}}\otimes\bra{\Phi_j^{(p)}}\bigr) \Tsqr
\bigl(\ket{\Phi_{i'}^{(n)}}\otimes\ket{\Phi_{j'}^{(p)}}\bigr)
\biggr]
\:,
\end{aligned}
}
where $\mathrm{Re}(z)$ denotes the real part of the complex number
$z$. Since $\Tsqr$ is a sum of one-body and two-body operators, 
the only contributions in the off-diagonal term of
Eq.(\ref{expval_Tsqr}) are therefore those for which
$\ket{\Phi_i^{(n)}}\otimes\ket{\Phi_j^{(p)}}$ and
$\ket{\Phi_{i'}^{(n)}}\otimes\ket{\Phi_{j'}^{(p)}}$ differ by 
a particle-hole excitation of order less than or equal to~2. 
In the following it will be useful to recall that, if
$\ket{\Phi_i^{(n)}}$ and $\ket{\Phi_j^{(p)}}$ are Slater determinants
of $N$ and $Z$ particles, respectively, then 
$\ket{\Phi_i^{(n)}}\otimes\ket{\Phi_j^{(p)}}$ is a Slater determinant
of $A=N+Z$ particles. Moreover, a Slater determinant $\ket{\Phi_i}$
without indication of its charge state $q$ is to be understood as a
product of a neutron $\ket{\Phi_j^{(n)}}$ and a proton
$\ket{\Phi_k^{(p)}}$ Slater determinants. Finally we recall that the
number of particles of each charge state $q$ is even since we treat
here even-even nuclei only.

Using the expressions for the isospin operator developed in Appendix~A,
it is easy to show that the diagonal matrix element of $\Tsqr$ can be
written in the form
\eq{
\label{n_T2_n}
\begin{aligned}
& \bigl(\bra{\Phi_i^{(n)}}\otimes\bra{\Phi_j^{(p)}}\bigr) \Tsqr
\bigl(\ket{\Phi_i^{(n)}}\otimes\ket{\Phi_j^{(p)}}\bigr)
= 
 \frac{A}{2} +\frac{(N-Z)^2}{4}\\
&  
-\sum_{k\in\Phi_i^{(n)}}\sum_{\ell\in\Phi_j^{(p)}}
|\overlap{k}{\ell}_{\rm space-spin}|^2\:,
\end{aligned}
}
where the notations $\overlap{k}{\ell}_{\rm space-spin}$ and
$\sum_{k\in\Phi_i^{(n)}}$ are defined in Appendix~A. 
It is important to note that the sums over the occupied
single-particle states of $\ket{\Phi_i^{(n)}}$ and
$\ket{\Phi_j^{(p)}}$ cannot be a priori reduced to sums over
time-reversed partner states, except for the many-body states 
(including $\ket{\Phi_0}$) in which the single-particle states 
are all paired. When this is not the case, the contributions 
of the form $\overlap{\overline{k}}{\ell}_{\rm space-spin}$ or 
$\overlap{k}{\overline{\ell}}_{\rm space-spin}$ (where 
$\ket{\overline{k}}$ is the time-reversed partner of $\ket{k}$) 
vanish. Therefore the expectation value of $\Tsqr$ in the
Hartree--Fock ground state $\ket{\Phi_0}$ is a special case of
Eq.~(\ref{n_T2_n}). The contribution 
$\elmx{\Psi}{\Tsqr}{\Psi}_{\rm diag}$ of the diagonal terms in 
Eq.~(\ref{n_T2_n}) to the expectation value of $\Tsqr$ finally writes
\eq{
\label{T2_diag}
\begin{aligned}
& \elmx{\Psi}{\Tsqr}{\Psi}_{\rm diag}=
\frac{A}{2}+\frac{(N-Z)^2}{4} \\
& -\sum_{i,j} \bigl(\chi_i^{(n)}\chi_j^{(p)}\bigr)^2
\sum_{k\in\Phi_i^{(n)}}\sum_{\ell\in\Phi_j^{(p)}}
|\overlap{k}{\ell}_{\rm space-spin}|^2\:.
\end{aligned}
}

To calculate the off-diagonal matrix elements we can exploit the fact
that one of the two Slater determinants of a given charge state is expressed 
as a 
$n$-particle--$n$-hole excitation with respect to the other
one. This gives simple expressions for the matrix elements but for
each pair of Slater determinants $\ket{\Phi_i^{(q)}}$ and
$\ket{\Phi_j^{(q)}}$ we have to determine the single-particle
states $\ket{i_1}$, ... , $\ket{i_n}$, $\ket{j_1}$, ... , $\ket{j_n}$
(hole or particle states of $\ket{\Phi_0}$) such that
$\ket{\Phi_j^{(q)}}=
\varphi_{ij}\,\ac{i_1}\cdots\ac{i_n}a_{j_1}\cdots 
a_{j_n}\ket{\Phi_i^{(q)}}\:,$
where $\varphi_{ij}=\pm 1$ is a phase factor determined in Appendix~B.

The non vanishing off-diagonal matrix element of $\Tsqr$ involving 
two Slater determinants differing by a 1-particle--1-hole excitation 
$\ac{i}a_j$ with $i\ne j$ is given by
\eq{
\label{n_T2_n1}
\begin{aligned}
& \bigl(\bra{\Phi_i^{(n)}}\otimes\bra{\Phi_j^{(p)}}\bigr) \Tsqr\ac{k} a_{\ell}
\bigl(\ket{\Phi_i^{(n)}}\otimes\ket{\Phi_j^{(p)}}\bigr) = \\
& -\delta_{kn}\delta_{\ell n}\sum_{m\in\Phi_j^{(p)}}
\overlap{\ell}{m}_{\rm space-spin}\,\overlap{m}{k}_{\rm space-spin} \\
& -\delta_{kp}\delta_{\ell p}\sum_{m\in\Phi_i^{(n)}}
\overlap{\ell}{m}_{\rm space-spin}\,\overlap{m}{k}_{\rm space-spin}\:.
\end{aligned}
}
Finally, for two Slater determinants differing by a 2-particle--2-hole 
excitation $\ac{i_1}\ac{i_2}a_{j_1}a_{j_2}$ with $\{i_1,i_2\}\cap
\{j_1,j_2\}=\emptyset$, we have
\eq{
\label{n_T2_n2}
\begin{aligned}
& \bigl(\bra{\Phi_i^{(n)}}\otimes\bra{\Phi_j^{(p)}}\bigr)
\Tsqr\ac{k_1}\,\ac{k_2}\,a_{\ell_1}a_{\ell_2}
\bigl(\ket{\Phi_i^{(n)}}\otimes\ket{\Phi_j^{(p)}}\bigr)=\\
& (\delta_{k_1 n} \,\delta_{k_2 p} \,\delta_{\ell_1 p} \,\delta_{\ell_2 n} 
 + \delta_{k_1 p} \,\delta_{k_2 n} \,\delta_{\ell_1 n} \,\delta_{\ell_2 p}) \times \\
& \overlap{\ell_1}{k_1}_{\rm space-spin}\,\overlap{\ell_2}{k_2}_{\rm space-spin} \\
&  - (\delta_{k_1 p} \,\delta_{k_2 n} \,\delta_{\ell_1 p} \,\delta_{\ell_2 n} 
   + \delta_{k_1 n} \,\delta_{k_2 p} \,\delta_{\ell_1 n} \,\delta_{\ell_2 p}) \times \\
& \overlap{\ell_1}{k_2}_{\rm space-spin}\,\overlap{\ell_2}{k_1}_{\rm space-spin}\:.
\end{aligned}
}

From Eqs.~(\ref{n_T2_n1}) and (\ref{n_T2_n2}) we deduce that the
non vanishing off-diagonal contribution 
$\elmx{\Psi}{\Tsqr}{\Psi}_{\rm off-diag}$ to the expectation value 
of $\Tsqr$ takes the form
\begin{widetext}
\eq{
\begin{aligned}
\label{T2_off-diag}
\elmx{\Psi}{\Tsqr}{\Psi}_{\rm off-diag} = &
2\sum_{i,j}\chi_i^{(n)}\bigl(\chi_j^{(p)}\bigr)^2
\sum_{i'=\mathrm{1p1h}(i)}\chi_{i'}^{(n)}
\bigl(\bra{\Phi_i^{(n)}}\otimes\bra{\Phi_j^{(p)}}\bigr) \Tsqr
\bigl(\ket{\Phi_{i'}^{(n)}}\otimes\ket{\Phi_j^{(p)}}\bigr) \\
& +2\sum_{i,j}\bigl(\chi_i^{(n)}\bigr)^2\chi_j^{(p)}
\sum_{j'=\mathrm{1p1h}(j)}\chi_{j'}^{(p)}
\bigl(\bra{\Phi_i^{(n)}}\otimes\bra{\Phi_j^{(p)}}\bigr) \Tsqr
\bigl(\ket{\Phi_i^{(n)}}\otimes\ket{\Phi_{j'}^{(p)}}\bigr) \\
& +2\sum_{i,j}\chi_i^{(n)}\chi_j^{(p)}
\sum_{\substack{i'=\mathrm{1p1h}(i)\\j'=\mathrm{1p1h}(j)}}
\chi_{i'}^{(n)}\chi_{j'}^{(p)}
\bigl(\bra{\Phi_i^{(n)}}\otimes\bra{\Phi_j^{(p)}}\bigr) \Tsqr
\bigl(\ket{\Phi_{i'}^{(n)}}\otimes\ket{\Phi_{j'}^{(p)}}\bigr)\:.
\end{aligned}
}
\end{widetext}
The first two terms of $\elmx{\Psi}{\Tsqr}{\Psi}_{\rm off-diag}$ are
calculated using respectively the first or the second term of the right 
hand side of Eq.~(\ref{n_T2_n1}), whereas the third term of
$\elmx{\Psi}{\Tsqr}{\Psi}_{\rm off-diag}$ corresponds to one of the
four series of $\delta$ products in the right hand side of 
Eq.~(\ref{n_T2_n2}). In practice, the sum of all off-diagonal terms is 
at least two orders of magnitude smaller than 
$\elmx{\Psi}{\Tsqr}{\Psi}_{\rm diag}$.

%
%

\subsection{Limiting cases}

We consider in this subsection two relevant limiting cases: the
Hartree--Fock limit and the limit of identical neutron and proton
single-particle states. We will refer to theses limits in Sect.~III to
interpret some results.

In the Hartree--Fock limit where $\chi_i^{(\tau)}=\delta_{i\,0}$, we
can deduce from Eqs.~(\ref{n_T2_n}), (\ref{n_T2_n1}) and
(\ref{n_T2_n2}) that the off-diagonal matrix elements 
vanish. The expectation value of $\Tsqr$ thus simply becomes
\eq{
\label{T2_HF}
\begin{aligned}
& \elmx{\Psi}{\Tsqr}{\Psi}= \elmx{\Phi_0}{\Tsqr}{\Phi_0}= \\
&\frac{A}{2}+\frac{(N-Z)^2}{4}
-\sum_{k\in\Phi_0^{(n)}}\sum_{\ell\in\Phi_0^{(p)}}
|\overlap{k}{\ell}_{\rm space-spin}|^2\:.
\end{aligned}
}

In the limit where the neutron and proton
single-particle states are assumed to be identical, the diagonal
contribution (\ref{T2_diag}) to 
$\elmx{\Psi}{\Tsqr}{\Psi}$ becomes
\eq{
\label{T2_n=p}
\begin{aligned}
\elmx{\Psi}{\Tsqr}{\Psi}_{\rm diag}= & T_0\left(T_0+1\right) \\
& +\sum_{\substack{i,j\\(i,j)\ne (0,0)}}(\chi_i^{(n)}\chi_j^{(p)})^2\,
\mathcal{R}(\Phi_i^{(n)},\Phi_j^{(p)})\:,
\end{aligned}
}
where $T_0=|T_z|$ and $\mathcal{R}(\Phi_i^{(n)},\Phi_j^{(p)})$ denotes
the relative excitation order of $\ket{\Phi_i^{(n)}}$ with respect of
$\ket{\Phi_j^{(p)}}$ (see Eq.~(\ref{reo}) of Appendix~B). Therefore
the isospin-mixing parameter takes, in this model case of 
identical neutron and proton single-particle states,
the simple form
\eq{
\label{alpha2_n=p}
\alpha^2=\frac{1}{2(T_0+1)}\,\sum_{\substack{i,j\\(i,j)\ne (0,0)}}(\chi_i^{(n)}\chi_j^{(p)})^2\mathcal{R}(\Phi_i^{(n)},\Phi_j^{(p)})\:.
}


\section{Results and discussion}
\label{res}

We study the isospin symmetry breaking through the isospin-mixing 
parameter $\alpha^2$ defined in Eq.~(\ref{alpha2}) 
for $T_z=-1$, $T_z=0$ and $T_z=1$ nuclei of eight elements, namely
$Z=12$ (Mg), $Z=16$ (S), $Z=20$ (Ca), $Z=24$ (Cr), $Z=28$ (Ni), $Z=36$
(Kr), $Z=40$ (Zr) and $Z=50$ (Sn).

To evaluate the expectation value $\elmx{\Psi}{\Tsqr}{\Psi}$ we need a
reliable description of the ground states of these nuclei. For that
purpose, we follow the two-step approach of
Ref.~\cite{Bonneau07_N=Z} where it was applied to study GS 
pairing properties of $N=Z$ nuclei in the mass $A\approx 70$ region. 
Since the above considered nuclei exhibit no triaxial deformation in their 
ground state, we can search for GS solutions possessing axial symmetry. 
In the first step, we determine the GS
deformation within the Hartree--Fock--BCS (HFBCS) approach. To do so,
we use the Skyrme interaction in its SIII parametrization~\cite{SIII}
in the mean-field channel, and the seniority force in the pairing
channel. For the latter we retain the same set of parameters as 
in Ref.~\cite{Bonneau07_N=Z}, where they were adjusted to reproduce 
experimental odd-even mass differences through a 3-point formula 
($G_0^{(n)}=17.70$~MeV, $G_0^{(p)}=15.93$~MeV, $\Delta\epsilon=6$~MeV 
and $\mu=0.2$~MeV). In practice, we use 15 oscillator major shells to 
expand the single-particle states on the cylindrical harmonic-oscillator 
basis and optimize the basis parameters at the GS deformation so as to 
obtain the lowest HFBCS binding energy. In the second step,
we calculate GS properties in the HTDA approach from the
above HFBCS solution. The residual interaction employed is the 
delta interaction of Ref.~\cite{Bonneau07_N=Z} adjusted in the same way 
as above for the seniority force but with $\Delta\epsilon=12$~MeV. 
The optimal values $V_0^{(q)}$ of the strength were found to be
$V_0^{(n)}=-340$~$\rm MeV.fm^3$ and $V_0^{(p)}=-306$~$\rm MeV.fm^3$ 
(this fit has been performed on the neutron pairing strength 
upon the simple approximation that $V_0^{(p)}$ is quenched by 
10\% with respect to $V_0^{(n)}$ because of the anti-pairing 
contribution of the Coulomb interaction).
However, keeping the same interaction strength throughout the whole 
considered nuclear region, we have taken care of the well-known 
$A^{-1/3}$ energy scale by varying the active pairing window: 
$\Delta\epsilon=12\,\times(72/A)^{1/3}$~MeV and 
$\mu=0.2\,\times(72/A)^{1/3}$~MeV, 
which yields for $A=72$ the same window parameters as those of 
Ref.~\cite{Bonneau07_N=Z}.

The GS properties calculated here are 
the charge radius $r_c$, the $\beta_2$ deformation parameter 
(see Appendix~C), the mass quadrupole ($Q_{20}$) and 
hexadecapole ($Q_{40}$) moments, the neutron and proton
pair-condensation energies $E_{cond}^{(q)}$, the trace of the operator
$\sqrt{\hat{\rho}(1-\hat{\rho})}$, which is equal
to the sum $\sum_i u_iv_i$ with $u_i=\sqrt{1-v_i^2}$ and
$v_i=\sqrt{\rho_{ii}}$, with $\hat{\rho}$ being the one-body density (see
Ref.~\cite{Bonneau07_N=Z}), and the total binding energy $E_b$. 
The results are reported in Table~\ref{tab_GS}.
\begin{table*}[t]
\caption{Ground-state properties of the twenty-one studied nuclei
  calculated within the HTDA approach. from left to right:  
  the charge radius $r_c$, the $\beta_2$ deformation parameter 
  (calculated as in Eq.~(\ref{eq_beta2}) of Appendix~C), the mass
  quadrupole ($Q_{20}$) and hexadecapole ($Q_{40}$) moments, the
  neutron and proton pair-condensation energies 
  $E_{\mathrm{cond}}^{(q)}$.\label{tab_GS}}
\begin{center}
\renewcommand{\arraystretch}{1.2}
\small
\begin{tabular}{cccccccccccccccc}
\hline\hline
 & \multirow{2}{*}{Nucleus} & \multirow{2}{*}{$r_c$ (fm)} & \multirow{2}{*}{$\beta_2$} & \multirow{2}{*}{$Q_{20}$ (fm$^2$)} & \multirow{2}{*}{$Q_{40}$ (fm$^4$)} & \multicolumn{2}{c}{$E_{\rm cond}$ (MeV)} & & \multicolumn{2}{c}{$\sum\limits_i u_iv_i$} & & \multirow{2}{*}{$E_{\rm b}$ (MeV)} & \\ 
 \cline{7-8} \cline{10-11} & & & & & & $n$ & $p$ & & $n$ & $p$ & & & \\ 
\hline
 & $^{22}$Mg & 3.107 & 0.346 & 92.1 & 147.8 & -0.737 & -0.687 & & 1.391 & 1.408 & & -169.038 & \\
 & $^{24}$Mg & 3.127 & 0.362 & 110.2 & 105.4 & -0.669 & -0.542 & & 1.232 & 1.133 & & -196.350 & \\
 & $^{26}$Mg & 3.085 & 0.220 & 69.4 & 42.4 & -0.715 & -0.574 & & 1.445 & 1.337 & & -215.591 & \\
 & $^{30}$S & 3.262 & 0.000 & -0.0 & 0.1 & -0.562 & -0.521 & & 1.191 & 1.385 & & -242.770 & \\
 & $^{32}$S & 3.299 & 0.192 & 83.3 & -35.2 & -0.604 & -0.488 & & 1.299 & 1.184 & & -268.322 & \\
 & $^{34}$S & 3.309 & 0.096 & 44.3 & -8.8 & -0.838 & -0.501 & & 2.027 & 1.352 & & -287.737 & \\
 & $^{38}$Ca & 3.479 & -0.002 & -1.0 & -0.0 & -0.855 & -0.774 & & 3.206 & 1.662 & & -311.842 & \\
 & $^{40}$Ca & 3.497 & 0.000 & 0.0 & 0.0 & -0.727 & -0.656 & & 1.488 & 1.516 & & -342.405 & \\
 & $^{42}$Ca & 3.510 & -0.003 & -1.6 & 1.8 & -1.578 & -0.530 & & 4.525 & 1.284 & & -361.798 & \\
 & $^{46}$Cr & 3.666 & 0.158 & 120.9 & 356.1 & -0.837 & -0.769 & & 2.242 & 2.087 & & -380.765 & \\
 & $^{48}$Cr & 3.709 & 0.241 & 203.7 & 666.9 & -0.670 & -0.571 & & 1.632 & 1.524 & & -409.845 & \\
 & $^{50}$Cr & 3.708 & 0.210 & 185.2 & 326.3 & -0.692 & -0.557 & & 1.714 & 1.581 & & -434.312 & \\
 & $^{54}$Ni & 3.788 & 0.002 & 2.0 & -0.4 & -1.058 & -0.393 & & 4.259 & 1.201 & & -452.236 & \\
 & $^{56}$Ni & 3.803 & 0.000 & -0.0 & 0.1 & -0.442 & -0.362 & & 1.254 & 1.148 & & -483.833 & \\
 & $^{58}$Ni & 3.828 & 0.002 & 2.3 & -0.3 & -0.652 & -0.353 & & 3.378 & 1.137 & & -502.966 & \\
 & $^{70}$Kr & 4.193 & -0.308 & -395.4 & 799.7 & -0.536 & -0.572 & & 1.701 & 1.863 & & -573.651 & \\
 & $^{72}$Kr & 4.222 & -0.352 & -468.1 & 1140.3 & -0.531 & -0.479 & & 1.670 & 1.647 & & -601.580 & \\
 & $^{74}$Kr & 4.235 & -0.350 & -487.0 & 1092.7 & -0.724 & -0.443 & & 2.658 & 1.581 & & -624.463 & \\
 & $^{78}$Zr & 4.392 & 0.392 & 785.9 & 1901.3 & -0.485 & -0.450 & & 1.594 & 1.614 & & -636.298 & \\
 & $^{80}$Zr & 4.414 & 0.398 & 834.2 & 1589.4 & -0.503 & -0.456 & & 1.663 & 1.606 & & -663.977 & \\
 & $^{82}$Zr & 4.439 & 0.418 & 919.0 & 2221.2 & -0.486 & -0.445 & & 1.753 & 1.586 & & -687.473 & \\
 & $^{98}$Sn & 4.524 & 0.000 & -0.7 & -10.6 & -0.841 & -0.367 & & 4.992 & 1.430 & & -793.287 & \\
 & $^{100}$Sn & 4.535 & 0.000 & 0.1 & -0.1 & -0.418 & -0.349 & & 1.498 & 1.392 & & -826.870 & \\
 & $^{102}$Sn & 4.554 & 0.000 & -0.1 & 7.6 & -0.709 & -0.342 & & 5.277 & 1.381 & & -846.359 & \\
\hline\hline
\end{tabular}
\end{center}
\end{table*}

\begin{figure}[h]
\includegraphics[width=0.45\textwidth]{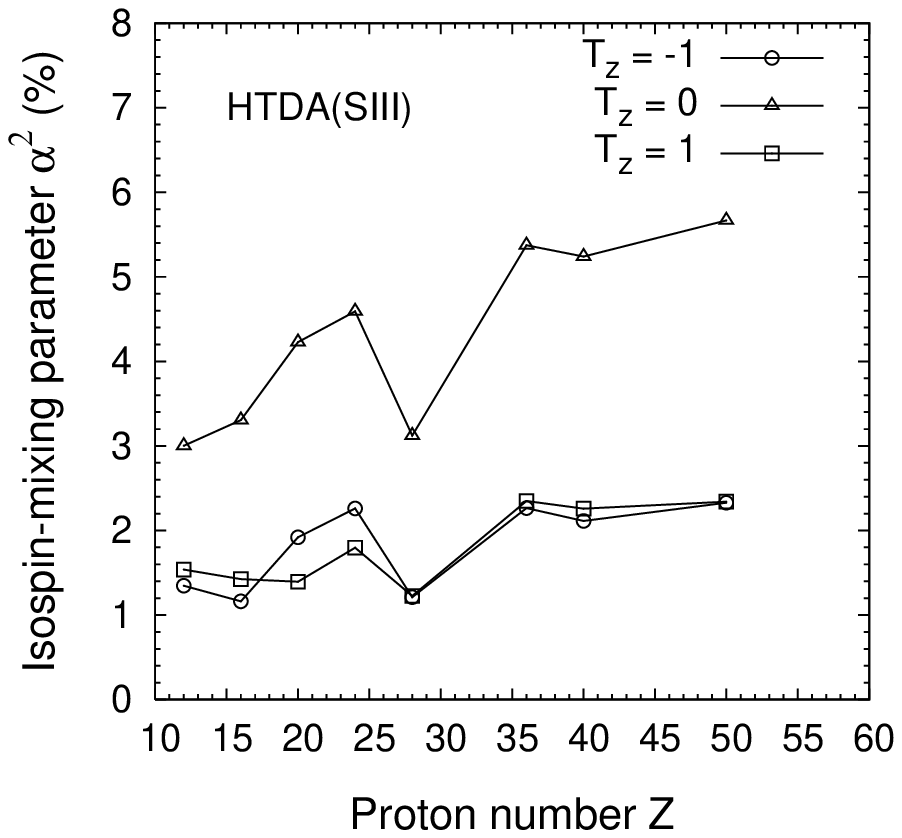}
\caption{\label{fig_alpha2}
Variation with $Z$ of the isospin-mixing parameter $\alpha^2$
calculated within the HTDA approach for $T_z=-1$ (open circles),
$T_z=0$ (open triangles) and $T_z=1$ (open squares) nuclei.
}
\end{figure}

The resulting HTDA ground state $\ket{\Psi}$ is then used to calculate
the expectation value of the $\Tsqr$ operator. In practice the 
off-diagonal term (\ref{T2_off-diag}) turns out to be negligible with
respect to the diagonal contribution (\ref{T2_diag}) and therefore can
be safely omitted in the calculations. Then, from the value of 
$\elmx{\Psi}{\Tsqr}{\Psi}$, we deduce the $T$-value defined by
\eq{
\label{T-value}
\elmx{\Psi}{\Tsqr}{\Psi}=T(T+1)\:.
}
We present the values of $\elmx{\Psi}{\Tsqr}{\Psi}$, $T$ and
$\alpha^2$ in the columns labeled ``HTDA'' in Table~\ref{tab_isospin} 
and in Fig.~\ref{fig_alpha2} we show the variation with $Z$ of the
isospin-mixing parameter $\alpha^2$ within the HTDA approach for the
above twenty four nuclei.
\begin{table*}[t]
\caption{Expectation value of $\Tsqr$, isospin $T$ from
Eq.~(\protect\ref{T-value}) and isospin-mixing parameter $\alpha^2$
from Eq.~(\protect\ref{alpha2}) calculated within the HTDA and HFBCS
approaches at the ground states determined in Table~\ref{tab_GS}. The
columns labeled ``HF'' correspond to the contributions to the above
three quantities coming from the Slater determinant $\ket{\Phi_0}$
in the HTDA ground state expansion of Eqs.~(\protect\ref{Psi_HTDA})
and (\protect\ref{Psi_tau_HTDA}). The values given in italic are
obtained without Coulomb interaction. \label{tab_isospin}}
\begin{center}
\renewcommand{\arraystretch}{1.2}
\small
\begin{tabular}{cccccccccccccccc}
\hline\hline
 & \multirow{2}{*}{} & \multirow{2}{*}{Nucleus} & & \multicolumn{3}{c}{$\langle\Psi|\hat{\mathbf{T}}^2|\Psi\rangle$} & & \multicolumn{3}{c}{$T$} & & \multicolumn{3}{c}{$\alpha^2$ (\%)} & \\ 
 \cline{5-7} \cline{9-11} \cline{13-15} & & & & HTDA & ``HF'' & HFBCS & &
 HTDA & ``HF'' & HFBCS & & HTDA & ``HF'' & HFBCS &  \\
\hline
 & \multirow{14}{*}{$T_z=-1$}
 & $^{22}$Mg & & 2.054 & 2.018 & 2.137 & & 1.018 & 1.006 & 1.045 & & 1.3 & 0.5 & 3.4 \\[-0.1cm]
 & & & & \it{2.049} & \it{2.014} & \it{2.186} & & \it{1.016} & \it{1.005} & \it{1.061} & & \it{1.2} & \it{0.3} & \it{4.6} \\
 &  & $^{30}$S & & 2.046 & 2.009 & 2.362 & & 1.015 & 1.003 & 1.116 & & 1.2 & 0.2 & 9.1 \\[-0.1cm]
 & & & & \it{2.038} & \it{2.003} & \it{2.270} & & \it{1.013} & \it{1.001} & \it{1.088} & & \it{0.9} & \it{0.1} & \it{6.8} \\
 &  & $^{38}$Ca & & 2.077 & 2.025 & 2.904 & & 1.025 & 1.008 & 1.276 & & 1.9 & 0.6 & 22.6 \\[-0.1cm]
 & & & & \it{2.052} & \it{2.003} & \it{2.890} & & \it{1.017} & \it{1.001} & \it{1.272} & & \it{1.3} & \it{0.1} & \it{22.3} \\
 &  & $^{46}$Cr & & 2.090 & 2.016 & 3.769 & & 1.030 & 1.005 & 1.505 & & 2.3 & 0.4 & 44.2 \\
 &  & $^{54}$Ni & & 2.048 & 2.019 & 2.857 & & 1.016 & 1.006 & 1.263 & & 1.2 & 0.5 & 21.4 \\[-0.1cm]
 & & & & \it{2.030} & \it{2.003} & \it{2.837} & & \it{1.010} & \it{1.001} & \it{1.257} & & \it{0.8} & \it{0.1} & \it{20.9} \\
 &  & $^{70}$Kr & & 2.091 & 2.038 & 3.967 & & 1.030 & 1.013 & 1.554 & & 2.3 & 1.0 & 49.2 \\[-0.1cm]
 & & & & \it{2.054} & \it{2.003} & \it{4.398} & & \it{1.018} & \it{1.001} & \it{1.656} & & \it{1.3} & \it{0.1} & \it{60.0} \\
 &  & $^{78}$Zr & & 2.085 & 2.043 & 2.680 & & 1.028 & 1.014 & 1.212 & & 2.1 & 1.1 & 17.0 \\[-0.1cm]
 & & & & \it{2.043} & \it{2.005} & \it{2.772} & & \it{1.014} & \it{1.002} & \it{1.238} & & \it{1.1} & \it{0.1} & \it{19.3} \\
 &  & $^{98}$Sn & & 2.093 & 2.068 & 2.896 & & 1.031 & 1.022 & 1.274 & & 2.3 & 1.7 & 22.4 \\[-0.1cm]
 & & & & \it{2.024} & \it{2.002} & \it{2.832} & & \it{1.008} & \it{1.001} & \it{1.256} & & \it{0.6} & \it{0.0} & \it{20.8} \\
\cline{2-15} & 
\multirow{14}{*}{$T_z=0$}
 & $^{24}$Mg & & 0.060 & 0.003 & 0.003 & & 0.057 & 0.003 & 0.003 & & 3.0 & 0.2 & 0.2 \\[-0.1cm]
 & & & & \it{0.056} & \it{0.000} & \it{0.000} & & \it{0.054} & \it{0.000} & \it{0.000} & & \it{2.8} & \it{0.0} & \it{0.0} \\
 &  & $^{32}$S & & 0.066 & 0.008 & 0.008 & & 0.062 & 0.007 & 0.007 & & 3.3 & 0.4 & 0.4 \\[-0.1cm]
 & & & & \it{0.057} & \it{0.000} & \it{0.000} & & \it{0.054} & \it{0.000} & \it{0.000} & & \it{2.8} & \it{0.0} & \it{0.0} \\
 &  & $^{40}$Ca & & 0.085 & 0.011 & 0.011 & & 0.078 & 0.011 & 0.011 & & 4.2 & 0.5 & 0.5 \\[-0.1cm]
 & & & & \it{0.070} & \it{0.000} & \it{0.000} & & \it{0.066} & \it{0.000} & \it{0.000} & & \it{3.5} & \it{0.0} & \it{0.0} \\
 &  & $^{48}$Cr & & 0.092 & 0.015 & 0.163 & & 0.085 & 0.015 & 0.142 & & 4.6 & 0.8 & 8.1 \\[-0.1cm]
 & & & & \it{0.081} & \it{0.000} & \it{0.545} & & \it{0.076} & \it{0.000} & \it{0.392} & & \it{4.1} & \it{0.0} & \it{27.3} \\
 &  & $^{56}$Ni & & 0.062 & 0.020 & 0.020 & & 0.059 & 0.020 & 0.020 & & 3.1 & 1.0 & 1.0 \\[-0.1cm]
 & & & & \it{0.040} & \it{0.000} & \it{0.000} & & \it{0.038} & \it{0.000} & \it{0.000} & & \it{2.0} & \it{0.0} & \it{0.0} \\
 &  & $^{72}$Kr & & 0.107 & 0.038 & 0.978 & & 0.098 & 0.037 & 0.608 & & 5.4 & 1.9 & 48.9 \\[-0.1cm]
 & & & & \it{0.068} & \it{0.000} & \it{1.667} & & \it{0.064} & \it{0.000} & \it{0.885} & & \it{3.4} & \it{0.0} & \it{83.4} \\
 &  & $^{80}$Zr & & 0.105 & 0.046 & 1.083 & & 0.096 & 0.044 & 0.655 & & 5.2 & 2.3 & 54.2 \\[-0.1cm]
 & & & & \it{0.051} & \it{0.000} & \it{1.091} & & \it{0.049} & \it{0.000} & \it{0.658} & & \it{2.6} & \it{0.0} & \it{54.6} \\
 &  & $^{100}$Sn & & 0.113 & 0.073 & 0.073 & & 0.103 & 0.069 & 0.069 & & 5.7 & 3.7 & 3.7 \\[-0.1cm]
 & & & & \it{0.036} & \it{0.000} & \it{0.000} & & \it{0.035} & \it{0.000} & \it{0.000} & & \it{1.8} & \it{0.0} & \it{0.0} \\
\cline{2-15} & 
\multirow{14}{*}{$T_z=1$}
 & $^{26}$Mg & & 2.062 & 2.025 & 2.189 & & 1.020 & 1.008 & 1.062 & & 1.5 & 0.6 & 4.7 \\[-0.1cm]
 & & & & \it{2.057} & \it{2.021} & \it{2.269} & & \it{1.019} & \it{1.007} & \it{1.087} & & \it{1.4} & \it{0.5} & \it{6.7} \\
 &  & $^{34}$S & & 2.057 & 2.003 & 3.493 & & 1.019 & 1.001 & 1.435 & & 1.4 & 0.1 & 37.3 \\[-0.1cm]
 & & & & \it{2.055} & \it{2.003} & \it{3.362} & & \it{1.018} & \it{1.001} & \it{1.401} & & \it{1.4} & \it{0.1} & \it{34.0} \\
 &  & $^{42}$Ca & & 2.056 & 2.018 & 3.037 & & 1.018 & 1.006 & 1.313 & & 1.4 & 0.4 & 25.9 \\[-0.1cm]
 & & & & \it{2.041} & \it{2.004} & \it{3.029} & & \it{1.013} & \it{1.001} & \it{1.311} & & \it{1.0} & \it{0.1} & \it{25.7} \\
 &  & $^{50}$Cr & & 2.072 & 2.026 & 2.528 & & 1.024 & 1.009 & 1.167 & & 1.8 & 0.7 & 13.2 \\[-0.1cm]
 & & & & \it{2.052} & \it{2.008} & \it{2.726} & & \it{1.017} & \it{1.003} & \it{1.225} & & \it{1.3} & \it{0.2} & \it{18.2} \\
 &  & $^{58}$Ni & & 2.049 & 2.018 & 3.193 & & 1.016 & 1.006 & 1.356 & & 1.2 & 0.5 & 29.8 \\[-0.1cm]
 & & & & \it{2.030} & \it{2.001} & \it{3.129} & & \it{1.010} & \it{1.000} & \it{1.338} & & \it{0.8} & \it{0.0} & \it{28.2} \\
 &  & $^{74}$Kr & & 2.094 & 2.042 & 3.910 & & 1.031 & 1.014 & 1.540 & & 2.3 & 1.1 & 47.7 \\
 &  & $^{82}$Zr & & 2.090 & 2.044 & 3.624 & & 1.030 & 1.015 & 1.468 & & 2.3 & 1.1 & 40.6 \\
 &  & $^{102}$Sn & & 2.094 & 2.069 & 3.582 & & 1.031 & 1.023 & 1.458 & & 2.3 & 1.7 & 39.5 \\[-0.1cm]
 & & & & \it{2.022} & \it{2.000} & \it{3.437} & & \it{1.007} & \it{1.000} & \it{1.420} & & \it{0.6} & \it{0.0} & \it{35.9} \\
\hline\hline
\end{tabular}
\end{center}
\end{table*}
Apart from a dip around $^{56}$Ni, $\alpha^2$ increases with $Z$,
faster for the $N=Z$ nuclei than for the others. Moreover the dip is
deeper for the former nuclei. We also note that the isospin-mixing
parameters for $T_z=-1$ and $T_z=1$ nuclei are very similar and are
about a factor of two smaller than for $T_z=0$ nuclei.

A precise determination of the isospin-mixing parameter requires that 
some great care be exerted in the calculations. This is illustrated 
in some typical examples in Appendix~D. Here we merely discuss the 
most important points.

First of all, we need to make sure that we have obtained a perfect 
consistency between the wave functions and the mean field including 
its Coulomb isospin-breaking part. In Appendix~D we show that a poor 
convergence of the iterative process may lead to drastic distortions 
in the isospin mixing evaluations.

A second important point is related to the quantal character of the 
assessed quantity. As exemplified in Appendix~D, it appears that in 
order to get reliable $\alpha^2$ values, one should include almost 
all Slater determinant components $\ket{\Phi_i}$ of the correlated 
wave function $\ket{\Psi}$, even those appearing in $\ket{\Psi}$ with 
a relatively minute probability $\bigl(\chi_i^{(q)}\bigr)^2$, because 
of constructive interference effects.

A last technical point is worth noting here. It deals with the question 
of the independence of our results with the harmonic oscillator basis 
parameters $b$ and $q$ (see their definition, e.g., in Ref.~\cite{FQKV}) 
in the expansion of the single-particle wave functions. One might have 
been concerned by the fact that this optimization has been performed 
merely at the level of the preliminary HFBCS calculations and not at 
the final stage of our HTDA approach. However it has been checked that 
an energy optimization of HTDA results leaves unchanged the calculated 
$\alpha^2$ values, as shown on one example in Appendix~D.

We now investigate several sources of isospin symmetry breaking.

\subsection{Roles of the neutron-proton mass difference and the
 Coulomb interaction}

We investigate separately the sensitivity of our results 
to the neutron-proton mass difference and to the presence of the
Coulomb interaction.

It turns out that the former plays virtually no role at all. For 
$^{40}$Ca, for instance, upon suppressing the one-body center of mass
correction (involving a $1/A$ term, ambiguous in this context), we
found that the mass difference is responsible for a variation in
$\alpha^2$ of the order of one part in $10^4$. It is neglected in the
remainder of the paper.

As can be seen from the columns ``HTDA'' of Table~\ref{tab_isospin}
where the results obtained without the Coulomb interaction are shown
in italic, the Coulomb interaction has a more noticeable effect which
still remains rather weak for the light nuclei considered here. In
this comparative study, we do not include the results for $^{46}$Cr,
$^{74}$Kr and $^{82}$Zr because the GS solutions without Coulomb
interaction differ too much from the ones obtained in the full
calculations to make a comparison meaningful. 

\subsection{Role of the particle-number conservation}

A very interesting issue consists in investigating the quality of the 
particle-number conserving pairing treatment (here in the $|T_z|=1$
channel only) obtained by the HTDA approach as compared to
approximations (as in the HFBCS calculations preliminary to 
our HTDA evaluation) which violate particle-number conservation. 
For that purpose, we evaluate the expectation value of $\Tsqr$ 
from the HFBCS ground state and deduce the value of the 
isospin-mixing parameter through Eq.~(\ref{alpha2}). The expectation value of 
$\Tsqr$ in a BCS state normalized to unity, noted $\ket{\rm BCS}$, 
reads
\eq{
\label{expval_T2}
\begin{aligned}
& \elmx{\rm BCS}{\Tsqr}{\rm BCS}= A+\frac{1}{4}\,(N-Z)^2-\sum_{i>0}v_i^4 \\
& -2\sum_{i>0}{}^{(n)}v_i^2\sum_{k>0}{}^{(p)}v_k^2\,
\bigl|\overlap{i}{k}_{\rm space-spin}\bigr|^2\:,
\end{aligned}
}
where the sums $\sum_{i>0}{}^{(n)}$, $\sum_{k>0}{}^{(p)}$ and 
$\sum_{i>0}$ run over neutron, proton and all pairs of time-reversed 
single-particle states of the form
$\bigl\{\ket{i},\ket{\bar{i}}\bigr\}$, respectively. The resulting
values of $\elmx{\rm BCS}{\Tsqr}{\rm BCS}$, $T$ and $\alpha^2$ are
reported in the columns labeled ``BCS'' of Table~\ref{tab_isospin} .

\begin{figure}[h]
\includegraphics[width=0.45\textwidth]{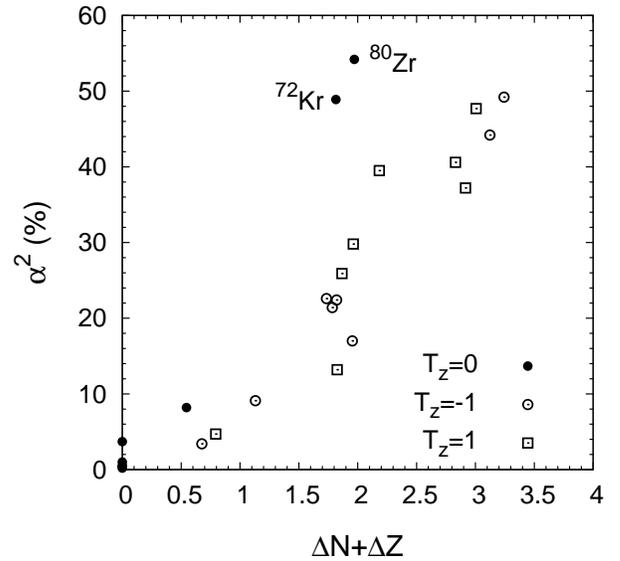}
\caption{\label{fig_alpha2_DeltaN}
Correlation between $\alpha^2$ and the particle-number fluctuation
$\Delta N+\Delta Z$ within the HFBCS approach for all twenty four
$T_z=-1$, $T_z=0$ and $T_z=1$ nuclei.}
\end{figure}

In cases where pairing correlations are ineffective in the BCS
treatment (one is then below the phase transition to the superfluid
phase), the value of $\alpha^2$ is very small, since in that
case essentially a single Slater determinant is describing the nuclear
state and the particle number is trivially conserved. In contrast, 
for those nuclei where pairing plays a non negligible role, the values
of the parameter $\alpha^2$ turn out to take on completely unrealistic
values as, e.g., for the nuclei $^{38}$Ca or $^{80}$Zr. In fact, 
as shown in Fig.~\ref{fig_alpha2_DeltaN}, there is a strong
correlation between $\alpha^2$ calculated in the HFBCS approach and
the particle-number fluctuation $\Delta N+\Delta Z$ in the BCS state, 
where $\Delta N_q$ is defined by 
\eq{
\Delta N_q=\sqrt{\elmx{\mathrm{BCS}}{\hat{N}_q^2}{\mathrm{BCS}}-N_q^2}\:.
}
With the exception of $^{72}$Kr and $^{80}$Zr,
all the points lie approximately on a straight line 
in the $(\alpha^2,\Delta N+\Delta Z)$ plane as can be seen in 
Fig.~\ref{fig_alpha2_DeltaN}.

\subsection{Role of the pairing correlations}

In order to assess the importance of pairing correlations on the
isospin-mixing rate, we also calculate the expectation value of 
$\Tsqr$ in the Slater determinant $\ket{\Phi_0}$ using
Eq.~(\ref{T2_HF}). The resulting values for
$\elmx{\Phi_0}{\Tsqr}{\Phi_0}$, $T$ and $\alpha^2$ are reported in the
columns labeled ``HF'' of Table~\ref{tab_isospin}. In general
$\elmx{\Phi_0}{\Tsqr}{\Phi_0}$ is different  from the value that would
result from a pure HF calculation because
$\ket{\Phi_0}$ is the Slater determinant built up from the
single-particle states resulting from the first-step HFBCS
calculation. This difference vanishes of course for nuclei in which
BCS predicts no pairing correlations, which is the case here for the
doubly magic nuclei as well as $^{24}$Mg and $^{32}$S.

\begin{figure}[h]
\includegraphics[width=0.45\textwidth]{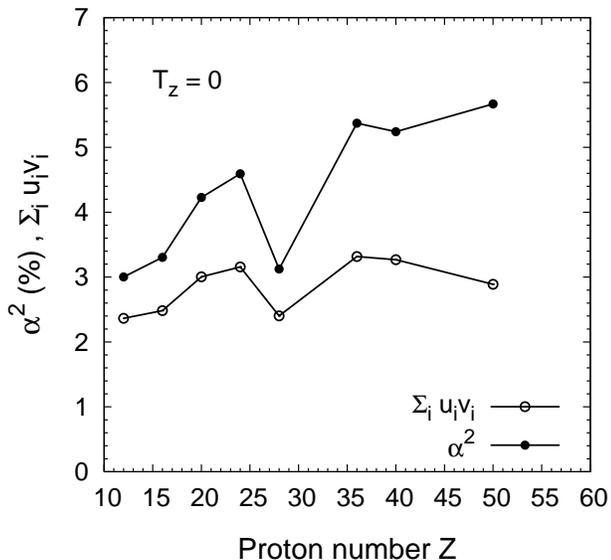}
\caption{\label{fig_alpha2_suv_Tz=0}
Correlated variations with $Z$ of $\alpha^2$ (solid circles) and $\sum_i
u_iv_i$ (open circles) for $T_z=0$ nuclei calculated within the HTDA
approach.
}
\end{figure}

For nuclei exhibiting weak pairing correlations, the
``HF'' results are, quite expectedly,
close to the HTDA predictions. Otherwise, the HTDA results are
significantly larger than the ``HF'' ones. The $T_z=0$ pairing
correlations are therefore an important source of isospin symmetry
breaking. This is conspicuous from Fig.~\ref{fig_alpha2_suv_Tz=0} which
shows the strong correlation between the variations of $\alpha^2$ and 
$\sum_i u_i v_i$ with $Z$ for $N=Z$ nuclei.

The large values of the isospin-mixing parameter found particularly in the
$T_z=0$ doubly-magic nuclei can be explained as follows taking the
example of $^{40}$Ca. As can be seen in Table~\ref{tab_isospin}, the
Coulomb contribution to the isospin mixing in HTDA calculations can be
considered as small. We can therefore make the approximation that the
neutron and proton single-particle states of $^{40}$Ca are virtually
identical and use Eq.~(\ref{alpha2_n=p}) to estimate
$\alpha^2$. Moreover, the largely dominant contributions to the
particle-hole excitation expansion of $\ket{\Psi}$ in
Eqs.~(\ref{Psi_HTDA}) and (\ref{Psi_tau_HTDA}) come from one-pair
excitations so that we can write $\alpha^2$ approximately as
\eq{
\begin{aligned}
\alpha^2\approx & \frac{1}{2(T_0+1)}\,\biggl[
\bigl(\chi_0^{(n)}\bigr)^2 \sum_{j\ne 0}\bigl(\chi_j^{(p)}\bigr)^2
\mathcal{R}(\Phi_0^{(n)},\Phi_j^{(p)})  \\
& +\bigl(\chi_0^{(p)}\bigr)^2 \sum_{i\ne 0}\bigl(\chi_i^{(n)}\bigr)^2
\mathcal{R}(\Phi_0^{(p)},\Phi_i^{(n)}) \\
& + \sum_{\substack{i\ne 0\\j\ne 0}}
\bigl(\chi_i^{(n)}\bigr)^2\bigl(\chi_j^{(p)}\bigr)^2
\mathcal{R}(\Phi_i^{(n)},\Phi_j^{(p)})
\biggr]\:,
\end{aligned}
}
where the relative excitation order of $\ket{\Phi_i^{(n)}}$ with
respect to $\Phi_j^{(p)}$ is simply given, here, by
\eq{
\label{reo_ij}
\mathcal{R}(\Phi_i^{(n)},\Phi_j^{(p)})=
\begin{cases}
0 & \mbox{if $i=0$, $j=0$;} \\
2 & \mbox{if $i=0$, $j\ne 0$ or $i\ne 0$, $j=0$;} \\
4 & \mbox{otherwise.}
\end{cases}
}
Since the states $\ket{\Psi^{(q)}}$ are normalized to unity, we have
\eq{
\sum_{i\ne 0}\bigl(\chi_i^{(q)}\bigr)^2=1-\bigl(\chi_0^{(q)}\bigr)^2\:,
}
and we finally obtain
\eq{
\label{alpha2_approx}
\alpha^2\approx \frac{1}{T_0+1}\biggl[2-\bigl(\chi_0^{(n)}\bigr)^2
-\bigl(\chi_0^{(p)}\bigr)^2
\biggr]\:.
}
Since Eq.~(\ref{alpha2_approx}) overestimates the importance of
one-pair excitations in $\ket{\Psi}$ through the estimates of
Eq.~(\ref{reo_ij}) and given the very small contribution to
$\ket{\Psi}$ coming from the particle-hole excitations other than
one-pair excitations, we conclude that the value of $\alpha^2$
calculated with Eq.~(\ref{alpha2_approx}) should lie between 
the values obtained in the full HTDA calculations without and with
Coulomb interaction. In the case of $^{40}$Ca, we find
$\bigl(\chi_0^{(n)}\bigr)^2=0.9804$ and
$\bigl(\chi_0^{(p)}\bigr)^2=0.9825$. This yields $\alpha^2\approx 
3.7\%$, which is slightly larger than the 3.5\% 
obtained in the HTDA
calculation without Coulomb interaction and smaller than the value of 4.2\%
from the full HTDA calculation, as expected.

Finally, it is interesting to note (see Table~\ref{tab_isospin}) that 
the value of $\alpha^2$ for a given nucleus obtained in a full HTDA 
calculation can be written, to a good approximation, as the sum of the
``no Coulomb'' HTDA result (including pairing correlations) and the ``HF'' 
result (no pairing, but including the full Coulomb field).

\subsection{Discussion}

\begin{figure}[h]
\includegraphics[width=0.45\textwidth]{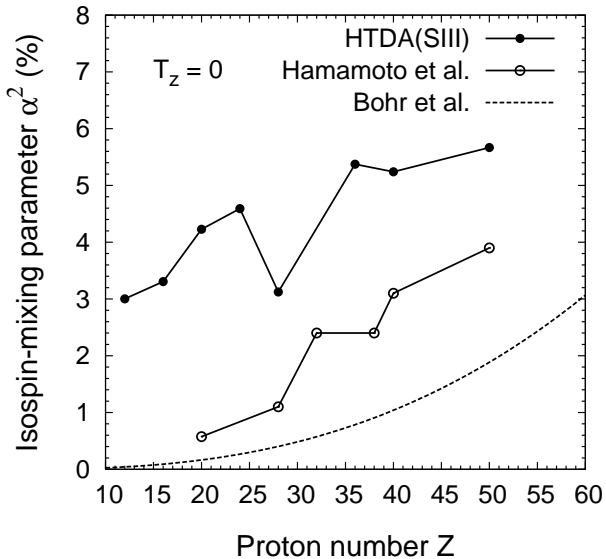}
\caption{\label{fig_alpha2_comp}
Variation with $Z$ of the isospin-mixing parameter $\alpha^2$ of
$T_z=0$ nuclei calculated within the HTDA approach (solid circles),
by Hamamoto \textit{et al.}~\protect\cite{sagawa} in the RPA
approach (open circles) and by Bohr \textit{et al.}~\protect\cite{BDM}
in a hydrodynamical model (dashed line). 
}
\end{figure}

In Fig.~\ref{fig_alpha2_comp} we compare, for the above eight $N=Z$ nuclei, 
the $\alpha^2$ values calculated in our HTDA model with the estimates 
obtained in the hydrodynamical model of Bohr, Damg{\aa}rd and 
Mottelson~\cite{BDM}, and with the calculations by Hamamoto and 
Sagawa~\cite{sagawa} in the Hartree--Fock-plus-RPA approach with the 
SIII Skyrme interaction. Each model predicts an
increasing trend of $\alpha^2$ with $Z$. The HTDA approach, as
presently applied with $|T_z|=1$ pairing correlations only, predicts a
larger isospin mixing than the RPA calculations (which do not
include pairing correlations).

It is important to recall that only $|T_z|=1$ pairing correlations
are considered here. We expect the values of $\alpha^2$ obtained
by including, in addition, pairing correlations in the $T_z=0$
channel to be smaller than the present values. Indeed, in presence of
proton-neutron correlations, the $|T_z|=1$ pairing correlations,
being effected by a smaller probability amplitude, would
contribute less to the total HTDA wave function. 
Correlatively, they would be replaced
essentially by configurations of the type
$\ac{i}a_{j}\ket{\Phi_0^{(n)}}\otimes\ac{k}a_{\ell}\ket{\Phi_0^{(p)}}$ where 
the neutron and proton hole states $\ket{j}$ and $\ket{\ell}$ on the
one hand, the neutron and proton particle states $\ket{i}$ and
$\ket{k}$ on the other hand, are similar. The relative excitation
order between two such one-particle--one-hole neutron,
one-particle--one-hole proton configurations would thus be on average 
smaller than that between two one-pair excitation neutron, one-pair
excitation proton configurations which reaches about 4 from
Eq.~(\ref{reo_ij}). However, the further addition of RPA correlations,
which can be treated in the HTDA framework on the same footing as the
pairing correlations, would compensate the effect of the $T_z=0$
pairing correlations and the net result may be close to the present
result.


\section{Conclusion}
\label{concl}

We have shown that the HTDA approach is a reliable model to address 
the isospin--mixing issue because such an approach can include the 
pairing correlations in a consistent way while conserving the 
particle-number, in contrast to Hartree--Fock--Bogoliubov and
Hartree--Fock--BCS treatments.

From HTDA calculations, we have learned that the difference between the
neutron and proton masses has a negligible impact on the expectation
value of $\Tsqr$ and that the effect of the Coulomb interaction is
rather small as compared to the effect of the $|T_z|=1$
correlations. Moreover the stronger isospin symmetry breaking is found
in the $N=Z$ nuclei.

To obtain a more complete description, both neutron-proton pairing and 
RPA-type correlations need to be taken into account. Both of these can be 
included in the HTDA framework in a consistent way. 
It is expected that these two 
types of correlations affect the present results with opposite signs.
Such a study is currently under way.


\section*{ACKNOWLEDGMENTS}

One of the authors (Ph. Q.) acknowledge the Theoretical Division at LANL 
for the excellent working conditions extended to him during numerous visits. 
This work has been supported by the U.S. Department of Energy under 
contract W-7405-ENG-36.


\setcounter{section}{0}
\renewcommand{\thesection}{APPENDIX \Alph{section}}

%
%

\section{Isospin operator $\Tsqr$ and one- and two-body matrix elements}

\setcounter{equation}{0}
\renewcommand{\theequation}{\Alph{section}-\arabic{equation}}

The operator $\Tsqr$ can be written as the sum
of a one-body operator $\hat{O}_1$ and a two-body operator $\hat{O}_2$
acting in the Fock space
\begin{gather}
\Tsqr=\hat{O}_1+\hat{O}_2\:, \\
\hat{O}_1=\sum_i\hat{\mathbf{t}}^2_i\:, \\
\hat{O}_2=\frac{1}{2}\sum_{i\ne j}
2\,(\hat{\mathbf{t}}_i\otimes\hat{\mathbf{t}}_j)\:,
\end{gather}
where $\hat{\mathbf{t}}_i\otimes\hat{\mathbf{t}}_j=
\hat{t}_x\otimes \hat{t}_x+\hat{t}_y\otimes \hat{t}_y+
\hat{t}_z\otimes \hat{t}_z$. We define the one-body and two-body 
operators $\hat{o}_1$ and $\hat{o}_2$ acting in the 
one-particle space and the two-particle space, respectively, by
\begin{gather}
\hat{o}_1=\hat{\mathbf{t}}^2\:, \\
\hat{o}_2=2\,(\hat{\mathbf{t}}\otimes\hat{\mathbf{t}})=
2\,\bigl(\hat{t}_x\otimes \hat{t}_x+\hat{t}_y\otimes \hat{t}_y+
\hat{t}_z\otimes \hat{t}_z\bigr)\:.
\end{gather}
Introducing the operators $\tp$ and $\tm$ defined by
\begin{gather}
\tp=\hat{t}_x+i\hat{t}_y\:, \\
\tm=\hat{t}_x-i\hat{t}_y\:,
\end{gather}
we can rewrite $\hat{o}_2$ as
\eq{
\label{o2}
\hat{o}_2=\tp\otimes\tm + \tm\otimes\tp + 2\,(\tz\otimes\tz)\:.
}

The Hartree--Fock basis is built up from the single-particle states 
generically noted $\ket{i}$. They describe either a neutron 
state or a proton state, so they are eigenstates of the isospin 
operators $\tsqr$ and $\tz$
\begin{gather}
\label{t2_i}
\tsqr\ket{i}=\frac{3}{4}\ket{i} \\
\tz\ket{i}=\tau_i\ket{i}=
\begin{cases}
\frac{1}{2}\,\ket{i} & \mbox{neutron} \\
-\frac{1}{2}\,\ket{i} & \mbox{proton}\:.
\end{cases}
\end{gather}
In practice, we expand the single-particle states
$\ket{i}$ on the cylindrical harmonic oscillator (HO) basis
$\bigl\{\ket{\alpha}\bigr\}$ as follows 
\eq{
\ket{i}=\sum_{\alpha} C_{\alpha}^{(i)} \ket{\alpha}\otimes\ket{t\,\tau_i}\:,
}
where $\alpha$ stands for the 4 quantum numbers $n_z$, $n_{\bot}$, $\Lambda$ 
(eigenvalue of $\hat{\ell}_z$) and $\Sigma$ (eigenvalue of
$\hat{s}_z$), $t=1/2$ and $\tau_i=\pm 1/2$ depending on the nature of
the particle. 

The action of the operators $\tm$ and $\tp$ on the single-particle states 
$\ket{i}$ is given by
\begin{gather}
\tm\ket{i}=\delta_{in}\sum_{\alpha} C_{\alpha}^{(i)} 
\ket{\alpha}\otimes\ket{t\,\tau_i-1}\:,\\
\tp\ket{i}=\delta_{ip}\sum_{\alpha} C_{\alpha}^{(i)} 
\ket{\alpha}\otimes\ket{t\,\tau_i+1}\:.
\end{gather}
The matrix elements of $\tm$ and $\tp$ in the Hartree-Fock basis thus
write
\begin{gather}
\label{i_Tm_k}
\elmx{i}{\tm}{k} = \delta_{ip}\delta_{kn}\bra{i}k\rangle_{\mathrm{space-spin}} 
\:,\\
\label{i_Tp_k}
\elmx{i}{\tp}{k} = \delta_{in}\delta_{kp}\bra{i}k\rangle_{\mathrm{space-spin}}
\:.
\end{gather}
In the above equations and elsewhere in this paper, the subscript
``space-spin'' attached to an overlap of single-particle states means
that the overlap is restricted to the space and spin variables
only. This allows to consider such overlaps between two
nucleonic states corresponding to different charges. 
Since the time reversal operator does not act on isospin, we have
\eq{
\elmx{i}{\hat{t}_{\pm}}{\bar{k}}=0\:,
}
where $\ket{\bar{k}}$ is the time-reversed conjugate state of $\ket{k}$, 
and
\eq{
\elmx{\bar{i}}{\hat{t}_{\pm}}{\bar{k}}=\elmx{i}{\hat{t}_{\pm}}{k}
\:.
}

From Eq.~(\ref{t2_i}) we easily get
\eq{
\label{elmx_o1}
\elmx{i}{\hat{o}_1}{j}=\frac{3}{4}\,\delta_{ij}
}
and using Eqs.~(\ref{o2}), (\ref{i_Tm_k}) and (\ref{i_Tp_k}), we can write 
the two-body matrix element $\elmx{ij}{\hat{o}_2}{k\ell}$ as
\eq{
\label{elmx_o2}
\begin{aligned}
\elmx{ij}{\hat{o}_2}{k\ell}= & 2\tau_i\tau_j\delta_{ik}\delta_{j\ell}
+(\delta_{ip}\delta_{kn}\delta_{jn}\delta_{\ell p}+
\delta_{in}\delta_{kp}\delta_{jp}\delta_{\ell n})\times \\
& \overlap{i}{k}_{\rm space-spin}\,
\overlap{j}{\ell}_{\rm space-spin}\:.
\end{aligned}
}

To close this appendix, we recall useful expressions for the expectation 
value of one-body and two-body operators in a Slater determinant
$\ket{\Phi_i}$ and related matrix elements. For a one-body operator
$\hat{O}_1$ we have
\begin{align}
& \elmx{\Phi_i}{\hat{O}_1}{\Phi_i}=
\sum_{k\in \Phi_i}\elmx{k}{\hat{o}_1}{k}\:,
\\
\label{n_O1_n1}
& \elmx{\Phi_i}{\hat{O}_1\ac{k} a_j}{\Phi_i}=\delta_j^h\delta_k^p\,
\elmx{j}{\hat{o}_1}{k}\:,
\end{align}
where the sum $\sum\limits_{k\in \Phi_i}$ runs over the
occupied single-particle states $\ket{k}$ of the Slater
determinant $\ket{\Phi_i}$. In Eq.~(\ref{n_O1_n1}) and below,
$\delta_j^h$ (resp. $\delta_k^p$) is equal to 1 if $\ket{j}$
(resp. $\ket{k}$) is a hole state (resp. particle state) with respect
to $\ket{\Phi_i}$ and 0 otherwise. For two-body operators we have
\begin{align}
& \elmx{\Phi_i}{\hat{O}_2}{\Phi_i}=
\frac{1}{2}\sum_{j,k\in\Phi_i}\elmx{jk}{\hat{o}_2}{\widetilde{jk}}\:,
\\
& \elmx{\Phi_i}{\hat{O}_2\ac{\ell} a_k}{\Phi_i}=
\delta_k^h\delta_{\ell}^p\,
\sum_{j\in\Phi_i} \elmx{j k}{\hat{o}_2}{\widetilde{j \ell}}\:, \\
& \elmx{\Phi_i}{\hat{O}_2\ac{k_1}\ac{k_2} a_{j_1}a_{j_2}}{\Phi_i}=
\delta_{j_1}^h\delta_{j_2}^h\delta_{k_1}^p\delta_{k_2}^p\,
\elmx{j_1j_2}{\hat{o}_2}{\widetilde{k_1k_2}}\:,
\end{align}
where $\ket{\widetilde{ij}}=\ket{ij}-\ket{ji}$.

%
%

\section{Comparison of two Slater determinants}

\setcounter{equation}{0}

Let us consider two Slater determinants $\ket{\Phi_i}$ and
$\ket{\Phi_j}$ built from the same set of orthonormal single-particle
basis states. They may therefore be thought of as 
$n$-particle--$n$-hole and $n'$-particle--$n'$-hole excitations on a
reference Slater determinant which may be chosen as the 
Hartree--Fock ground state $\ket{\Phi_0}$
\begin{gather}
\ket{\Phi_i} =
\ac{\beta_1}\cdots\ac{\beta_n}a_{b_1}\cdots a_{b_n}\,\ket{\Phi_0}\:, \\
\ket{\Phi_j} =
\ac{\gamma_1}\cdots\ac{\gamma_{n'}}a_{c_1}\cdots a_{c_{n'}}\,\ket{\Phi_0}\:,
\end{gather}
where $\beta_1 < \cdots < \beta_n$ and $\gamma_1 < \cdots <
\gamma_{n'}$ are two sets of particle states (with respect to
$\ket{\Phi_0}$) and $b_1 < \cdots < b_n$ and $c_1 < \cdots < c_{n'}$
are two sets of hole states. We can express $\ket{\Phi_j}$ as a
function of $\ket{\Phi_i}$ as
\eq{
\ket{\Phi_j} =
\ac{\gamma_1}\cdots\ac{\gamma_{n'}}a_{c_1}\cdots a_{c_{n'}}
\ac{b_n}\cdots\ac{b_1}a_{\beta_n}\cdots a_{\beta_1}
\,\ket{\Phi_i}\:.
}
We denote by $\mathcal{H}$ the set of hole states in common between 
$\bigl\{b_1,\cdots,b_n\bigr\}$ and $\bigl\{c_1,\cdots,c_{n'}\bigr\}$
\eq{
\mathcal{H}=\bigl\{c_{h_i'}=b_{h_i},1\leqslant i \leqslant N_h\bigr\}\:,
}
where $N_h$ is the number of hole states in common. Similarly
$\mathcal{P}$ is the set of particle states in common between 
$\bigl\{\beta_1,\cdots,\beta_n\bigr\}$ and
$\bigl\{\gamma_1,\cdots,\gamma_{n'}\bigr\}$ 
\eq{
\mathcal{P}=\bigl\{\gamma_{p_i'}=\beta_{p_i},1\leqslant i \leqslant
N_p\bigr\}\:,
}
where $N_p$ is the number of hole states in common. Therefore it can
be shown that
\eq{
\label{comp_Slaters}
\begin{aligned}
& \ket{\Phi_j} = \\
& \varphi_{ij}
\biggl(\prod_{\substack{k=1\\ \gamma_k\notin\mathcal{P}}}^{n'}
\ac{\gamma_k}\biggr)
\biggl(\prod_{\substack{k=n\\ b_k\notin\mathcal{H}}}^1\ac{b_k}\biggr)
\biggl(\prod_{\substack{k=1\\ c_k\notin\mathcal{H}}}^{n'} a_{c_k}\biggr)
\biggl(\prod_{\substack{k=n\\ \beta_k\notin\mathcal{P}}}^1 a_{\beta_k}\biggr)
\ket{\Phi_i}\:,
\end{aligned}
}
where the associated relative phase is given by
\eq{
\varphi_{ij}=(-1)^{n+n'+N_h+\sum\limits_{k=1}^{N_h}(h_i-h_i')+
\sum\limits_{k=1}^{N_p}(p_i-p_i')}\:.
}
Changing the order of the creation and/or annihilation operators in
Eq.~(\ref{comp_Slaters}) would change the sign of $\varphi_{ij}$. 

Finally the relative excitation order $\mathcal{R}(\Phi_i,\Phi_j)$
between the two Slater determinants $\ket{\Phi_i}$ and $\ket{\Phi_j}$, 
defined as the number of creation (or annihilation) operators in
Eq.~(\ref{comp_Slaters}), is simply given by 
\eq{
\label{reo}
\mathcal{R}(\Phi_i,\Phi_j)=n+n'-(N_h+N_p)\:.
}

%
%

\section{Nuclear shape and size quantities}

\setcounter{equation}{0}

Starting from the nuclear shape determined in a self-consistent way
by the HTDA solution, we can extract a quadrupole deformation
parameter $\beta_2$ by approximating the nuclear shape by the
equivalent spheroid having the same root-mean-square mass radius 
$r_m$ and mass quadrupole moment $Q_{20}$ as the actual nucleus. 
The semi-axes $c$ (along the symmetry axis) and $a$ (in the 
perpendicular direction) are related to $r_m$ and $Q_{20}$ through
\begin{gather}
A\,r_m^2=\int d^3\mathbf{r}\,\rho(\mathbf{r})\,\mathbf{r}^2 =
\frac{1}{5}(2a^2+c^2)\:, \\
Q_{20}=2\int
d^3\mathbf{r}\,\rho(\mathbf{r})\,\mathbf{r}^2P_2(\cos\theta) =
\frac{2}{5}\,A\,(c^2-a^2)\:,
\end{gather}
where $A=N+Z$, $\rho(\mathbf{r})$ is the isoscalar nuclear density
(sum of neutron and proton contributions) and $P_2$ is the Legendre
polynomial of degree 2. 

The $\beta_2$ parameter is then calculated for this equivalent 
spheroid by expanding the nuclear radius in polar coordinates 
according to the $\beta_l$-parametrization~\cite{Moller95}
\begin{align}
R(\theta)&=\frac{a}{\sqrt{1-\alpha\,\cos^2 \theta}}\\
&=R_0\,\left(1+\sum_{l=1}^{\infty}\beta_l\,Y_l^0(\theta)\right)\:,
\end{align}
with
\eq{
\alpha=1-\frac{a^2}{c^2}\:.
}
This allows us to derive the analytical expression of $\beta_2$ for 
the equivalent spheroid as a function of $\alpha$ as
\eq{
\label{eq_beta2}
\beta_2=\begin{cases}
\sqrt{5\pi}\left[\frac{3}{2\alpha}\left(1-\frac{\sqrt{\alpha(1-\alpha)}}{{\rm Arcsin}\, \sqrt{\alpha}}\right)-1\right] & \alpha\in]0;1[ \\
0 & \alpha=0 \\
\sqrt{5\pi}\left[\frac{3}{2\alpha}\left(1-\frac{\sqrt{-\alpha(1-\alpha)}}{{\rm ln}\, (\sqrt{-\alpha}+\sqrt{1-\alpha})}\right)-1\right] & \alpha<0
\end{cases}\:.
}

As for the mass hexadecapole moment $Q_{40}$, we calculate it 
using the following expression with usual notation
\eq{
Q_{40} =\int d^3\mathbf{r}\,\rho(\mathbf{r})\,r^4\,Y_4^0(\theta)\:.
}

Finally, the charge radius $r_c$ is calculated as in Refs.~\cite{Negele,Vautherin72} 
through
\eq{
\label{rc}
r_c^2=\int d^3\mathbf{r}\,\int d^3\mathbf{s}\,f_p(\mathbf{r}-\mathbf{s})\,
\rho_p(\mathbf{r})\,\mathbf{r}^2\:,
}
where $\rho_p(\mathbf{r})$ is the proton density and $f_p(\mathbf{x})$ 
denotes the proton form factor. With a Gaussian form for the latter, 
$f_p(\mathbf{x})=\exp(-\mathbf{x}^2/r_0^{\,2})/(r_0\,\sqrt{\pi})$, we 
have
\eq{
r_c^2=r_p^2+\frac{3}{2}\,r_0^{\,2}\:,
}
with
\eq{
r_p^2=\int d^3\mathbf{r}\,\rho_p(\mathbf{r})\,\mathbf{r}^2\:.
}
In our calculations we choose to use the value $r_0=0.65$~fm 
($\frac{3}{2}r_0^{\, 2}=0.64$~fm$^2$) from Ref.~\cite{Vautherin72}.

%
%

\section{Technical aspects of the calculations}

\setcounter{equation}{0}

We illustrate in this appendix the importance of several technical 
aspects of the mean-field calculations that can have a substantial 
impact on the quality of the results for the isospin-mixing parameter 
$\alpha^2$.

\begin{figure}[h]
\begin{center}
\includegraphics[width=0.45\textwidth]{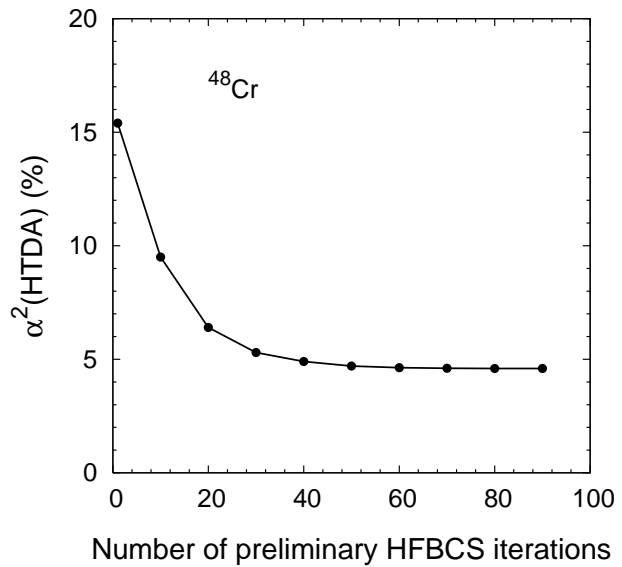}
\end{center}
\caption{Variation of the isospin-mixing parameter $\alpha^2$
  calculated within the HTDA approach with the number of preliminary
  HFBCS iterations, in the case of $^{48}$Cr.
  \label{fig_alpha2_iterHFBCS}}
\end{figure}

First, the choice of the reference Slater determinant $\ket{\Phi_0}$
for HTDA calculations is very
important. Figure~\ref{fig_alpha2_iterHFBCS} shows the variation of
the isospin-mixing parameter $\alpha^2$ calculated within the HTDA
approach with the number of preliminary HFBCS iterations, in the case
of $^{48}$Cr. As an initial potential we choose the Woods--Saxon
potential including a spin-orbit term with the same parameters for
neutron and protons (hence without Coulomb interaction for the first
iteration).
From the decreasing and saturating trend obtained here, we
conclude that the consistency of the underlying mean-field from which
$\ket{\Phi_0}$ is determined plays an important role. In other words
a poor mean-field is a source of spurious isospin symmetry breaking.

\begin{figure}[h]
\begin{center}
\includegraphics[width=0.45\textwidth]{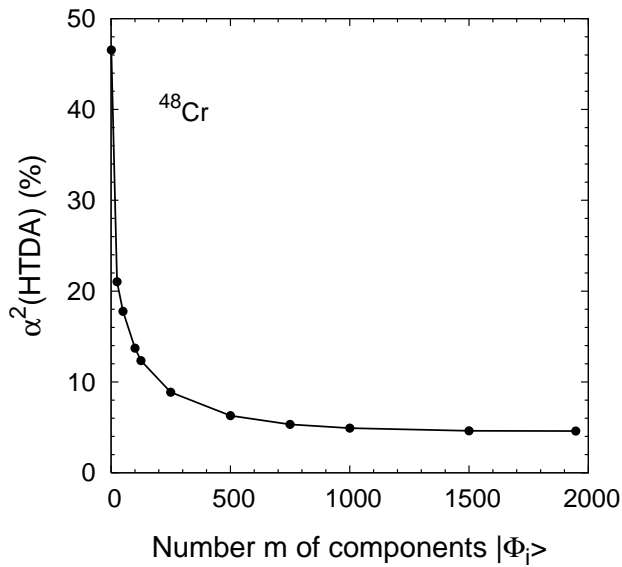}
\end{center}
\caption{Variation of $\alpha^2$ with the number of
  Slater determinants $\ket{\Phi_i}$, from the expansion of the HTDA
  ground state $\ket{\Psi}$, that are used to calculate
  $\alpha^2$.
\label{fig_alpha2_maxph}} 
\end{figure}

\begin{figure}[h]
\begin{center}
\includegraphics[width=0.45\textwidth]{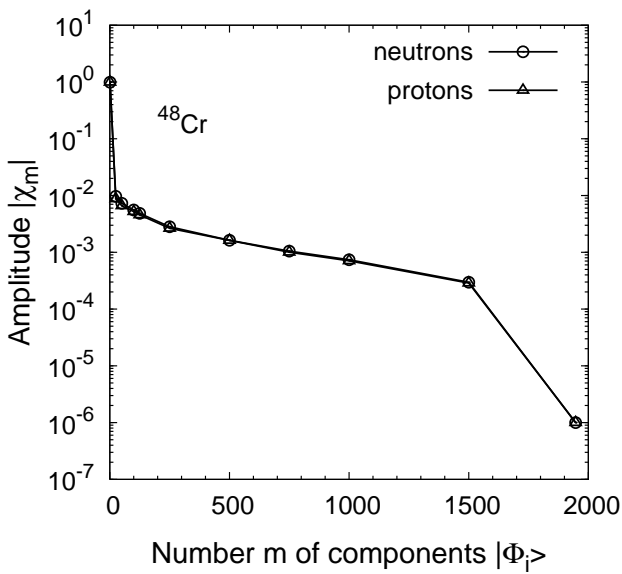}
\end{center}
\caption{Variation of the neutron (open circles) and proton (open 
triangles) Slater determinant amplitudes $|\chi_m|$ with 
the number $m$ of Slater determinants retained from the expansion 
of $\ket{\Psi}$ for the calculation of $\alpha^2$. The first point 
is for $m=1$, the dominant component in $\ket{\Psi}$.
\label{fig_chi_maxph}} 
\end{figure}

The second aspect of importance is the convergence of $\alpha^2$ with
the number $m$ of Slater determinants contributing to $\ket{\Psi}$ 
retained in the calculation of $\elmx{\Psi}{\Tsqr}{\Psi}$. The Slater
determinants $\ket{\Phi_i^{(q)}}$ entering the expansion
(\ref{Psi_tau_HTDA}) of $\ket{\Psi^{(q)}}$ are arranged in decreasing
order of $|\chi_i^{(q)}|$. The results are shown in 
Fig.~\ref{fig_alpha2_maxph}, whereas in Fig.~\ref{fig_chi_maxph} 
we present the variation of the neutron and proton Slater 
determinant amplitudes $|\chi_m|$ with $m$.  

Finally, we have checked for three nuclei that the values of the
isospin-mixing parameter calculated in the HTDA approach is not
sensitive to the procedure of optimization of the harmonic oscillator
basis parameters $b$ and $q$ (with the notation of Ref.~\cite{FQKV}),
namely at the level of the preliminary HFBCS calculations or at the
final stage of our HTDA approach. The optimized values of $b$ and
$q$ obtained in each of the two schemes are reported in 
Table~\ref{tab_opt_bq} together with the resulting values of $\alpha^2$ 
for $^{24}$Mg, $^{40}$Ca and $^{80}$Zr.
We recall that the cylindrical harmonic oscillator 
basis used in all calculations contains 15 major shells, which corresponds 
to $N_0=14$ in the notation of Ref.~\cite{FQKV}.
\begin{table}[h]
\caption{Optimized values of $b$ and $q$ obtained in the two
  optimization schemes (at the level of the preliminary HFBCS
  calculations or at the final stage of the HTDA calculations) and
  resulting values of the isospin-mixing parameter $\alpha^2$
  calculated in the HTDA approach for $^{24}$Mg, $^{40}$Ca and
  $^{80}$Zr.\label{tab_opt_bq}}
\begin{tabular}{cccccccccc}
\hline\hline
\multirow{2}{*}{Nucleus} & & \multicolumn{3}{c}{HFBCS optimization} &
& \multicolumn{3}{c}{HTDA optimization} & \\
\cline{3-5} \cline{7-9}
 & & $b$ & $q$ & $\alpha^2$ (\%) & & $b$ & $q$ & $\alpha^2$ (\%) & \\
\hline
 $^{24}$Mg & & 0.65 & 1.28 & 3.001 & & 0.65 & 1.05 & 2.999 & \\
 $^{40}$Ca & & 0.66 & 1.00 & 4.228 & & 0.66 & 1.00 & 4.228 & \\
 $^{80}$Zr & & 0.60 & 1.37 & 5.229 & & 0.58 & 1.45 & 5.264 & \\
\hline\hline
\end{tabular}
\end{table}


\end{document}